\documentclass[twocolumn,showpacs,amsmath,amssymb]{revtex4}
\input epsf
\usepackage{appendix}
\usepackage{rotating}
\begin{document}
\title {Finite temperature analysis of a quasi2D dipolar gas}
\author{Christopher Ticknor}
\affiliation{Theoretical Division, Los Alamos National Laboratory, Los Alamos, 
New Mexico 87545, USA}
\date{\today}
\begin{abstract}
We present finite temperature analysis of a quasi2D dipolar gas. To do this,
we use the Hartree Fock Bogoliubov method within the Popov approximation.
This formalism is a set of non-local equations containing the
dipole-dipole interaction and the condensate and thermal correlation
functions, which are solved self-consistently.
We detail the numerical method used to implement the scheme.
We present density profiles for a finite temperature dipolar gas in quasi2D, and  
compare these results to a gas with zero-range interactions.
Additionally, we analyze the excitation spectrum and study the impact of the
thermal exchange. 
\end{abstract} 
\maketitle
\section{Introduction}
Two hot topics in ultracold physics are two dimensional (2D) gases 
and dipolar gases. 
Reduced dimensionality enhances the quantum mechanical character of 
the system.  In a homogeneous 2D system, quantum phase fluctuations are 
so strong that at finite temperature phase coherence cannot be established 
and condensation does not occur.
There is, however, an intriguing phase transition called the 
Berezinskii-Kosterlitz-Thouless (BKT) transition \cite{BKT} which occurs when 
the temperature is lowered and there is no longer enough thermal energy to 
unbind vortex and anti-vortex pairs. This binding of vortices reduces
phase fluctuations so that quasi-long range order can form \cite{BKTrev}.
Interestingly, a trapped 2D gas can Bose Einstein condense (BEC) at finite 
temperature \cite{BKTrev,trap}.
The BKT and BEC transition have been studied in trapped ultracold gases,
first by observations of phase defects \cite{zoran},
direct observation of vortices\cite{JILA}, and then 
changes in the density profile due to the onset of superfluidity \cite{NIST}.
Additionally, universality has been observed near the 
BKT transition \cite{chicago}. 
Successful methods which study such systems are Monte Carlo \cite{critical,boris},
mean field \cite{HolzmannMF,bissetMF} and classical-field methods \cite{simula,foster,bisset}.

In dipolar systems the interactions are non-local and the possibility of 
creating strongly correlated gases are tantalizing, especially in 
quasi-2D (q2D) where zero point motion in the axial direction is included.  
For example, some studies have used 
Monte Carlo methods to study phase transitions such as 
crystallization \cite{buchler}. This study included    
temperature as a variable and observed melting of the dipolar crystal.  
The BKT transition of a homogeneous dipolar gas has been studied with Monte 
Carlo methods to examine superfluid fraction and excitation spectra \cite{dBKT}.
Other studies have looked at the structure of the dipolar gas
and the impact of temperature on the phase of the gas \cite{Nho,jain}.
Such studies have focused on strongly interacting dipolar systems.
However, the bulk of the work on q2D gases has been at zero temperature. 
For example,  phonon instability \cite{nath} and 
anisotropic superfluidity \cite{aniso} have been predicted.

Recently, chromium (Cr) and dysprosium (Dy) have been Bose 
condensed and exhibited strong dipolar effects \cite{Cr,Dy}.
Additionally, progress towards a q2D dipolar gas has been made with a 
layered dipolar system for Cr atoms in a one dimensional optical 
lattice \cite{layer}.  With such progress on dipolar gases, 
direct study of q2D dipolar systems has begun.
An important question is how will dipolar interactions impact the 
quantum behavior of a q2D gas? 
The long range nature of the dipolar interaction may lead to 
interesting physics for the thermal system, especially relating the BKT 
transition and phase coherence.  
Additionally, little work studying the impact of temperature on q2D dipolar gases
has been conducted for reasonable experimental parameters.

With an eye toward this unexplored physics and aiding experiments, we
study the finite temperature physics of q2D trapped dipolar gases.
We use the standard beyond mean-field method: the Hartree Fock Bogoliubov
within the Popov approximation (HFBP) \cite{fetter}.  
The HFBP has been successful in studying ultracold atoms, for example
it has been used to explain the temperature dependence of collective 
excitations in a 3D BEC \cite{temp-mode}.
In the case of 3D dipolar gases, Ref. \cite{ronen} used the HFBP
to study temperature effects on the biconcave structure dipolar gases 
can have \cite{roton}.  This study neglected the thermal exchange.
Other recent work used mean field methods to study the 
stability of finite temperature dipolar gases \cite{mftemp}.
  
The aim of this paper is two fold.
First, we develop a numerical method to solve the non-local HFBP.
This method includes the non-local interaction and exchange effects.
We show in detail how the interaction and thermal exchange effects are included;
it relies on a parallel implementation of the method.
The method can also be applied to dipolar fermions.
Second, we compare the contact gas and the dipolar gas with the HFBP.
In the comparison of the two gases, we look at 
the condensate number as a function of temperature.
We also articulate the role of thermal exchange in the gas by 
solving the system with and without thermal exchange.   
We  compare the density profiles of the gas at various temperatures.  
We look at the excitation spectrum as a function 
of both temperature and total particle number. We also classify the lowest 
excitation modes. To the author's knowledge, these results are the first for a 
q2D trapped-finite temperature dipolar gas with thermal exchange effects.
\section{Equations of motion}
We study a quasi-2D dipolar system at finite temperature.
To do this, we will employ the Hartree Fock Bogoliubov method
within the Popov approximation with non-local 
interactions \cite{fetter,griffin}.
The HFB breaks the second quantized bosonic field operator
into a condensate and non-condensate (thermal) component:
$\hat\Psi=[\sqrt{N_0}\phi_0(x)+\hat\theta(x)]$ where we have replaced 
$\hat a_0\rightarrow\sqrt{N_0}$ and $N_0$ is the number of atoms in the condensate mode. 
Here $x$ represents all required coordinates, for the 2D case it will 
be $\vec\rho$. We will use the Bogoliubov transformation for the thermal 
part: $\hat\theta(x)=\sum_\alpha [u_\alpha(x)\hat a_\alpha e^{-i\omega_\alpha t}
-v_\alpha^*(x)\hat a^*_\alpha e^{i\omega_\alpha t}]$ where $\hat a$ ($\hat a^*$) is 
the bosonic annihilation (creation) operator for a quasi-particle (hole).  
They obey the bosonic commutation relations: 
$[\hat a_\alpha,\hat a_\beta^*]=\delta_{\alpha\beta}$ and  
$[\hat a_\alpha,\hat a_\beta]=[\hat a_\alpha^*,\hat a_\beta^*]=0$.
To derive the equations of motion we start with the 
Heisenberg equation of motion with a non-local interaction \cite{griffin}.
We find for our system the modified non-local Gross Pitaevskii equation (GPE) is
\begin{widetext}
\begin{eqnarray}
\mu \phi_0(x)=&&\left(H_0+\int dx^\prime n(x^\prime) V\right)\phi_0 (x)
+\int dx^\prime [\tilde n(x,x^\prime)
-\tilde m(x,x^\prime)]V\phi_0 (x^\prime),
\label{GPE}
\end{eqnarray}
where $H_0$ is the kinetic energy and trapping potential. 
The interaction potential, $V(x-x^\prime)$, has been written as $V$ for simplicity.
The total density,  
$n(x)=n_0(x)+\tilde n(x)$ is made up of condensate density and thermal density.
The non-local correlation functions are:
$n_0(x,x^\prime)=N_0\phi_0^*(x) \phi_0(x^\prime)$ and $n_0(x)=n_0(x,x)$;
$\tilde n(x,x^\prime)=\langle \hat \theta^*(x) \theta(x^\prime)\rangle$ is the thermal 
correlation function and the local thermal density is $\tilde n(x)=\tilde n(x,x)$.
$\tilde m(x,x^\prime)=\langle \hat \theta(x) \hat\theta(x^\prime)\rangle$ is the anomalous 
thermal correlation function, and in the Popov approximation is neglected.
The quasi-particle wavefunctions are required to construct the thermal 
and anomalous correlation functions.  At equilibrium, these wavefunctions can be
found from a non-local Bogoliubov-de Gennes equation:
\begin{eqnarray}
\omega_\alpha u_\alpha(x)=&&\left(H_0-\mu+\int dx^\prime n(x^\prime)V\right)u_\alpha(x)
+\int dx^\prime [n_0(x,x^\prime)+\tilde n(x,x^\prime)]Vu_\alpha (x^\prime)\label{uveq}
\\\nonumber
&&+\int dx^\prime \tilde m(x,x^\prime)V v_\alpha (x^\prime)
-\int dx^\prime m_0(x,x^\prime) V v_\alpha(x^\prime) 
\\
-\omega_\alpha v_\alpha = &&\left(H_0-\mu+\int dx^\prime n(x^\prime)V\right)v_\alpha(x) 
+\int dx^\prime [n_0(x,x^\prime)+\tilde n(x,x^\prime)]Vv_\alpha (x^\prime)
\nonumber\\ \nonumber
&&+\int dx^\prime \tilde m^*(x,x^\prime)Vu_\alpha (x^\prime)
-\int dx^\prime m_0^*(x,x^\prime)  Vu_\alpha(x^\prime).
\end{eqnarray}
\end{widetext}
The total number of atoms is $N=N_0+\tilde N$ and $\tilde N=\int dx \tilde n(x)$. 
The normalization of the quasi-particle wavefunction is 
$1=\int dx (|u_\alpha|^2-|v_\alpha|^2)$.  The anomalous condensate correlation function is 
$m_0(x,x^\prime)=N_0\phi(x)\phi(x^\prime)$.
The thermal correlation functions are defined in terms of the quasi-particles:
\begin{widetext}
\begin{eqnarray}
&&\tilde n(x,x^\prime)=\label{nxxp}\sum_\alpha\left[u_\alpha^*(x^\prime)u_\alpha(x)
+v_\alpha(x^\prime)v_\alpha^*(x)\right]N_{be}^\alpha+v_\alpha(x^\prime)v_\alpha^*(x)
\\&&\tilde m(x,x^\prime)=\label{mxxp}\sum_\alpha\left[u_\alpha(x^\prime)v^*_\alpha(x)
+v^*_\alpha(x^\prime)u_\alpha(x)\right]N_{be}^\alpha+u_\alpha(x^\prime)v^*_\alpha(x),
\end{eqnarray}
\end{widetext}
where $N_{be}^\alpha=(Ze^{\hbar\omega_\alpha/kT}-1)^{-1}=\langle \hat
a_\alpha^* \hat a_\alpha\rangle$ where $k$ is Boltzmann's constant, $T$ is the temperature,
and $Z=1+1/N_0$.  This relation is simply the Bosonic occupation of thermal modes.
These equations are consistent with Refs. \cite{ronen}, \cite{griffin},  
and \cite{esry} when the appropriate simplification is made.

To make the discussion of  Eqs. (\ref{GPE}) and (\ref{uveq}) simpler,
we introduce a compact notation:
\begin{eqnarray}
&&(H_0+D+\tilde X+\tilde M)\phi_0=\mu\phi_0 \label{gp1},\\
&&(H_0-\mu+D+X)u_\alpha+Mv_\alpha=\omega_\alpha u_\alpha \label{uv},\\
&&(H_0-\mu+D+X)v_\alpha+Mu_\alpha=-\omega_\alpha v_\alpha \nonumber.
\end{eqnarray}
The first term is the standard kinetic and potential energy.
For the remaining interaction terms, 
we use the capital letters to indicate the inclusion of the
$x^\prime$ integral, so
the  direct interaction is $D=D_0+\tilde D$ (condensate and
thermal), and for example $D_0(x)=\int dx^\prime [n_0(x^\prime)]V(x-x^\prime)$.
The exchange interaction is $X=X_0+\tilde X$ (condensate and thermal),
and for example
 $\tilde X(x,x^\prime)=\int dx^\prime [\tilde n(x,x^\prime)]V(x-x^\prime)$.
Finally, the anomalous correlation function is $M=\tilde M-M_0$, and for example
 $M(x,x^\prime)=\int dx^\prime [\tilde m(x,x^\prime)
 -m_0(x,x^\prime)]V(x-x^\prime)$.
We will use the Popov approximation, which assumes $\tilde M=0$, leading 
to $M=-X_0$ if $\phi_0$ is real. This approximation keeps the spectrum gapless.

We project out the condensate mode from the quasi-particles, as was done in
Refs. \cite{castin1,morgan} which uses $Q=1-|0\rangle\langle0|$ and 
$\langle x |0\rangle=\phi_0(x)$.  
The projection operator is applied to all operators that are not
in $H_{GP}=H_0-\mu+D+\tilde X$, such as $M_0$ and $X_0$, and for example, 
$M_0\rightarrow QM_0Q$.  
This projection method keeps the spectrum gapless.

\section{The Interaction}
We are interested in the q2D dipolar system assuming no axis of symmetry, so
we take the dipole moment to be $\vec d=d[\hat z \cos(\alpha)+\hat x
\sin(\alpha)]$, and
we assume harmonic confinement and that $\omega_z\gg\omega_\rho$ where the $\omega_i$
are the trapping frequencies ( $\omega_x=\omega_y=\omega_\rho$).  This allows us to 
assume only one transverse mode is occupied in the $z$ direction.
We can then evaluate the interaction by factoring the wavefunction as
$\psi(\vec r)=\chi_0(z)\psi(\vec\rho)$, where $\psi$ is either $\phi_0$, $u_\alpha$, 
or $u_\alpha$ and $\chi_0$ is eigenstate of $H_0$ in the $z$ direction. 
We are then able to integrate out $z$, and
obtain an effective interaction.  We will also consider a contact 
interaction. The interaction we use is:
\begin{eqnarray}
V(\vec \rho) =g\delta(\vec\rho) +g_d V_{dd}(\vec\rho),
\end{eqnarray}
where 
$g=\sqrt{8\pi} \hbar^2a_s/m l_z$ is the strength of the contact 
interaction, $a_s$ is s-wave the scattering length ($a_s\ll l_z$), 
$l_z=\sqrt{\hbar/m \omega_z }$ is the axial harmonic 
oscillator length, $g_d=d^2/\sqrt{2\pi} l_z$ and $d$ is the dipole moment of 
the particles.  This interaction leads to the dipolar length scale: 
$l_d=md^2/\hbar^2$.  In practice, we rescale the equations of motion into 
oscillator units, where the units of energy and length are: $\hbar\omega_\rho$ and 
$l_\rho=\sqrt{\hbar/m\omega_\rho}$.  After this rescaling, the dipolar interaction strength
becomes  $g_d=l_d/(l_z\sqrt{2\pi})$.  Finally, the dipolar interaction in q2D is:
\begin{eqnarray}
&&V_{dd}(\vec \rho-\vec \rho^\prime)=\int dz \chi_0^2(z) 
\int dz^\prime  \chi_0^2(z^\prime) V_d^{3D}(\vec r-\vec r^\prime)\label{ddi3D}
\\\nonumber
&&V_{dd}^{3D}(\vec r)=d^2(1-3(\hat d \cdot \hat r)^2)/r^3.
\end{eqnarray}
The interaction is evaluated in momentum space.  As an example, the
direct interaction is:
\begin{eqnarray}
&&D(\vec \rho)=\mathcal{F}^{-1}\left[{n}(\vec{k}) 
\tilde V\left(\vec{k}\right)\right]
\\\nonumber
&&\tilde V(\vec{k})=g_d \frac{4\pi}{3}F\left(\frac{\vec{k} l_z}{\sqrt{2}}\right),
\end{eqnarray}
where $\mathcal{F}$ is the Fourier transform operator 
and ${n}(\vec{k}) = \mathcal{F}[n(\vec{\rho})]$.  
$\tilde V_k$ is the momentum representation of the interaction.
The function $F(\vec k)$ 
is the $k$-space dipolar interaction for the q2D geometry. It has two 
contributions coming from polarization perpendicular to and in the direction 
of the dipole tilt, 
$F(\vec k)=\cos^2(\alpha)F_{\perp}(\vec k)+\sin^2(\alpha)F_{\parallel}(\vec k)$ 
where $\alpha$ is the angle between $\hat{z}$ and the polarization vector 
$\hat d$.  Its contributions
are $F_{\parallel}(\vec k)=-1+3\sqrt{\pi}(k_d^2/k)e^{k^2}\mbox{erfc}(k)$, 
where $\vec k_d$ is the wave vector along the polarization direction in 
the x-y plane, erfc is the complementary error 
function and $F_{\perp}(\vec k)=2-3\sqrt{\pi}ke^{k^2}\mbox{erfc}(k)$ \cite{uwe,nath,aniso}.

To evaluate the exchange interaction, we use a simple, yet memory intensive, 
solution: we explicitly construct the interaction on the same space-space 
grid as the correlation functions.  We put $n(\vec\rho,\vec\rho{}^\prime)$ on 
a grid specified by the indices $s$ and $t$, representing $\vec\rho$ and 
$\vec\rho^\prime$.  Each grid has $n_s$ spatial points, and a correlation 
function has $n_s^2$ grid points.

To evaluate the interaction, we work in momentum space. The interaction on the 
space-space grid is: 
\begin{eqnarray}
V_{st}=W^T_{sk}\tilde V_kW_{kt},
\end{eqnarray}
where $W_{ks}$ is the operator that transforms between space ($s$) and 
k-space ($k$) representation via the spectral basis set \cite{dpgp}.  
The operator takes the place of the Fourier operator in our algorithm.
An important aspect of the method is that it 
regularizes the interaction and avoids the logarithmic divergence as $\vec \rho\rightarrow\vec\rho^\prime$ 
encountered by directly evaluating Eq. (\ref{ddi3D}) because 
$W_{kt}$ involves a projecting onto the Gauss-Hermite basis set \cite{dpgp}. 
Constructing the interaction involves the most costly step, 
requiring $n_s^3$ 
operations. This is performed only once for a set of interaction
parameters ($l_z$ and $\alpha$) and basis set size. 

We have constructed a parallel implementation of the method.  
We distribute the $s$ index ($\vec\rho$ grid) across the computing 
nodes ($p$); we will denote this distributed index as $s_p$.  This way, each core 
handles only a fraction of the correlation functions and interaction.
In the future, we will construct the interaction in the partial wave expansion:
$V(\vec \rho)=\sum_m V_m(\rho,\phi)$. 
This expansion will be necessary to handle an  interaction which
depends on quantum numbers such as $n_z$ (confinement in $z$) or spin.

\section{numerical implementation}
We solve Eqs. (\ref{gp1}) and (\ref{uv}) by expanding the wavefunctions on 
the basis 
that diagonalizes $H_0$.  We will focus on the harmonically trapped case where 
$H_0 \chi_\alpha=\epsilon_\alpha \chi_\alpha$, $\int d\vec\rho \chi_\alpha^*(\vec\rho)\chi_\beta(\vec\rho)=\delta_{\alpha\beta}$, and $\epsilon_\alpha=\hbar\omega_\rho( m_x+m_y+1)$ where $m_\alpha$ is an integer.   
In a basis set, the evaluation of the direct interaction term, 
$D_{\alpha\beta}$, is straight forward:
$\int d\vec\rho \chi_\alpha(\vec\rho) [\int dx^\prime V(\vec\rho-\vec\rho^\prime) n(\vec\rho^\prime)]\chi_\beta(\vec\rho)$, where
the quantity in the square brackets is the effective potential from the 
interaction. The more complicated terms are the non-local exchange terms, 
such as the thermal exchange: $\tilde X_{\alpha\beta}$ which is 
$\int d\vec\rho \chi_\alpha^*(\vec\rho)\int d\vec\rho^\prime \tilde n(\vec\rho,\vec\rho^\prime)V(\vec\rho-\vec\rho^\prime)\chi_\beta(\vec\rho^\prime)$. 

The non-local exchange term in the spectral basis is: 
$X_{\alpha\sigma}= U_{\alpha s}^T V_{st} n_{st} U_{t\sigma}$, where
$U_{s\sigma}$ is the transformation between the spectral basis ($\sigma$) and 
space basis ($s$). The numerical procedure to evaluate this on node $p$ is:
(1) $Y_{s_pt}=V_{s_pt}n_{s_pt}$, multiply the interaction and correlation function;
(2) $h_{s_p \sigma}=Y_{s_pt}U_{t\sigma}$, project the $t$ spatial basis onto the 
$\sigma$ spectral basis;
(3) $X^p_{\alpha\sigma}=\sum_{s_p}U^T_{s_p\alpha }h_{s_p\sigma}$, project $s_p$ spatial basis
onto the  $\alpha$ spectral basis; and 
(4) $X_{\alpha\sigma}=\sum_{p}X^p_{\alpha\sigma}$, collect and sum. 
In comparison, the direct terms are straight forward and evaluated in momentum
space at each step.  $\Phi_s=W_{sk}\tilde V_{k}W_{kt}n_{t}$, where $\Phi_s$ is the 
effective potential.  Then we must project onto the basis: 
$D_{\alpha\sigma}= U_{\alpha s}^T \Phi_{s} U_{s\sigma}$, and this is done in parallel.

The procedure to find the full solution is: 
set temperature, $T$, and $N_0$ and pick a targeted value for the total 
number of particles, $N_{target}$, and set $\tilde n=0$ and $\tilde m=0$:
\begin{enumerate}
\item Solve Eq. (\ref{GPE}) for $\phi_0$ and $\mu$ (in basis set).
\item Construct condensate exchange term, $X_0$, with $n_0(\vec\rho,\vec\rho^\prime)$ 
and put in basis set.
\item Solve Eq. (\ref{uveq})  for $u_\alpha$ and $v_\alpha$.
\item Construct $\tilde n(\vec\rho,\vec\rho^\prime)$ and 
$\tilde m(\vec\rho,\vec\rho^\prime)$.
\item Use the semi-classical LDA to supplement the thermal tail (see below),
which gives $\tilde N$ and therefore $N=N_0+\tilde N$.
\item Construct thermal and anomalous exchange terms, $\tilde X$ and 
$\tilde M$ (if not using Popov approximation), and put in basis set representation.
\item Adjust $N_0$ to get the desired $N_{target}$ of atoms. We require N to 
be within 1$\%$ of $N_{target}$.
\item Go back to step 1 until self-consistency is reached.  We converge the 
number of thermal atoms, so that $|\tilde n_i-\tilde n_{i-1}|<5\cdot10^{-5}$, 
when we are near enough $N_{target}$.
\end{enumerate}
With the numerical method in hand, we are ready to proceed to the physical 
examples and the results.  

The semi-classical, local density approximation (LDA) is very 
important for large $\vec\rho$ and high momentum states on the grid.
At high temperature with a manageable grid size (much bigger than the 
condensate), we find that there can be a appreciable number of atoms outside 
of quadrature grid.  For the large $\vec\rho$ contributions, the density is 
low enough that the analytic, 
non-interacting solution can be used to account for these off-grid particles.  
We integrate the analytic solution over a large region outside of the grid. 
We have tested that this does not impact the convergence of the end result.
We include the impact of higher transverse modes in the 
same analytic manner on and off the grid.
We also assume that these particles in the higher $z$ modes are non-interacting.
This is reasonable, especially because symmetry prevents the first excited 
transverse mode from colliding with the condensate with $m=0$ symmetry.
Additionally, a noticable number of thermal particles are in momentum states 
beyond calculated quantum  spectrum  \cite{reidl,bergman}. To include these
thermal atoms we use the method from Ref. \cite{ronen}.

\section{Physical examples}
We will study Cr and Dy atoms.  These atoms 
have been Bose condensed and shown to exhibit strong dipolar effects \cite{Cr,Dy}. 
The first example we choose is: $2\cdot10^3$ Cr atoms 
with  $l_d=45a_0$ where $a_0$ is the Bohr radius.  This gives $g_d=0.0028$ and $g=0$
in trap units. 
Additionally, we pick $l_z/l_\rho=0.1$, this is from using 
$(\omega_\rho,\omega_z)$=$2\pi$ (16,1600) Hz.  
We have found that our numerical 
procedure using 465 basis states with an energy cut of 30$\hbar \omega_\rho$ converges.
To compare this to a contact gas, we match the chemical potential at zero 
temperature by varying $g$.
We will only consider $\alpha=0$ here, that is perpendicular polarization where
the dipolar interaction is isotropic and repulsive.
We also study Dy with a  trapping potential of $(\omega_\rho,\omega_z)$=$2\pi$ 
(10,1000) Hz.  With $l_d=400a_0$, this gives $g_d=0.034$ in trap units.  
This is strongly interacting and to maintain the 
Kohn mode (center of mass slosh mode), we have used a low atom number, and
for our example we use 300 atoms \cite{Dy}.
It is important to note that the Cr example is very similar to recent 
experiments \cite{layer}. In that work, they were able to vary $l_z/l_\rho$ between 
0.1 and 0.17 and have up to 2000 atoms in a layer \cite{layer}. 
However, this is a layered system and interlayer dipolar interactions are important. 

In contrast to a homogenous system, a trapped ideal 2D gas can have
Bose condensation \cite{trap}. In this case, the critical temperature is at $T_{c0}/\hbar\omega=\sqrt{6N}/\pi\sim0.78\sqrt{N}$.  
We will use this temperature to rescale our findings so that, to first order, we 
remove the number dependence of the results. This is important because, as we vary the 
temperature, we do not have exactly the same number of particles between every 
calculation. In the thermodynamic limit, the population of the condensate is 
$N_0/N=1-(T/T_{c0})^2$ \cite{trap}.

For reference, we give several specific numerical examples in  Table \ref{table}.
For two temperatures, we report $\mu$, $N_0/N$, $N$, and the Kohn mode energy (should be 1$\hbar\omega_\rho$).
For the Dy ($g_d=0.034$) example, at $T/T_{c0}=0.5$, the Kohn mode has noticably 
deviated from 1 by about 2.5$\%$, this is about the worst the convergence
of this mode gets as a function of temperature. 
For both dipolar examples, we show the results for when  
thermal exchange is negelected ($\tilde X=0$).
The lower temperature solutions are similar to the GPE solutions at $T=0$.

\begin{table}
\caption{Numerical examples of the HFBP for the q2D dipolar gas and contact gas.}
\begin{tabular}{|c|c|c|c|c|c|}
\hline
&$T/T_{c0}$&$\mu/\hbar\omega$ & $N_0/N$&$N$&Kohn Mode$/\hbar\omega$\\
\hline
$g_d$=0.0028&0.05&3.822&0.994&2004&0.9992\\
&0.5&3.457&0.658&1994&0.995\\
\hline
$g_d$=0.0028&0.05&3.814&0.994&2004&0.9996\\
${\tilde X=0}$&0.5&3.341&0.660&1994&0.997\\
\hline
$g$=0.021&0.05&3.829&0.993&2004&0.9991\\
&0.5&3.541&0.662&1993&0.9995\\
\hline
$g_d$=0.034&0.05&5.124&0.983&300&0.991\\
&0.5&4.919&0.525&298&0.976\\
\hline
$g_d$=0.034&0.05&5.094&0.983&300&0.996\\
${\tilde X=0}$&0.5&4.524&0.5500&303&0.987\\
\hline
\label{table}
\end{tabular}
\end{table}

\section{Properties of a q2D dipolar gas at  finite temperature}
\begin{figure}
\includegraphics[width=80mm]{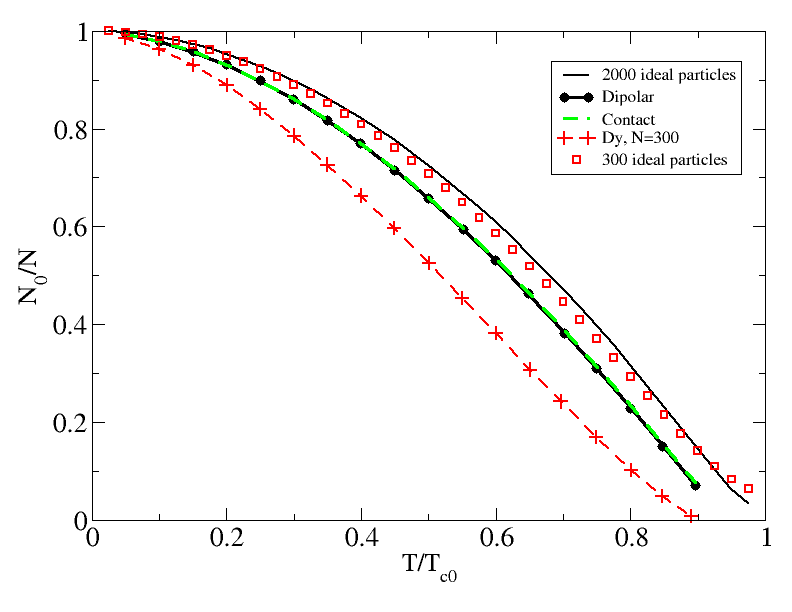}
\caption{$N_0/N$ as a function of Temperature for the Cr
dipolar gas (thick black line with circles), the contact gas (green dash dot), and
for 2000 ideal trapped particles (solid black line).  Additionally,
a gas of 300 Dy particles (dashed red with +) and 300 ideal particles are
shown (red open squares).
}
\label{ncomp}
\end{figure}
In this section we study the properties of a q2D dipolar gas at  finite temperature,
and compare it to a contact gas and a dipolar system without the thermal exchange.
We look at several different aspects of these systems.
First, we look at the condensate number as a function of temperature.
Then, we compare the interaction contributions to the total energy.
Third, we compare the density profiles of the contact and dipolar gas, 
at various temperatures.  Fourth, we look at the impact of including the
thermal exchange on the density profile of the gas.
Finally, we look at the excitation spectrum as a function of temperature 
and total particle number. We also classify the lowest excitation modes. 

Figure \ref{ncomp} shows the condensate fraction for the Cr
dipolar gas (black line with circles), the contact gas (green dash dot line),
 for 2000 ideal particles (solid black line),
300 Dy particles (dashed red), and 300 ideal particles (red open squares).
The condensate fraction for the dipolar gas and contact gas are shifted 
down from the ideal gas (black line), with the dipolar gas being slightly lower than 
the contact gas.  For the Dy example, it is clear that the interaction strongly 
depletes the condensate mode and lowers the critical temperature.
\begin{figure}
\includegraphics[width=85mm]{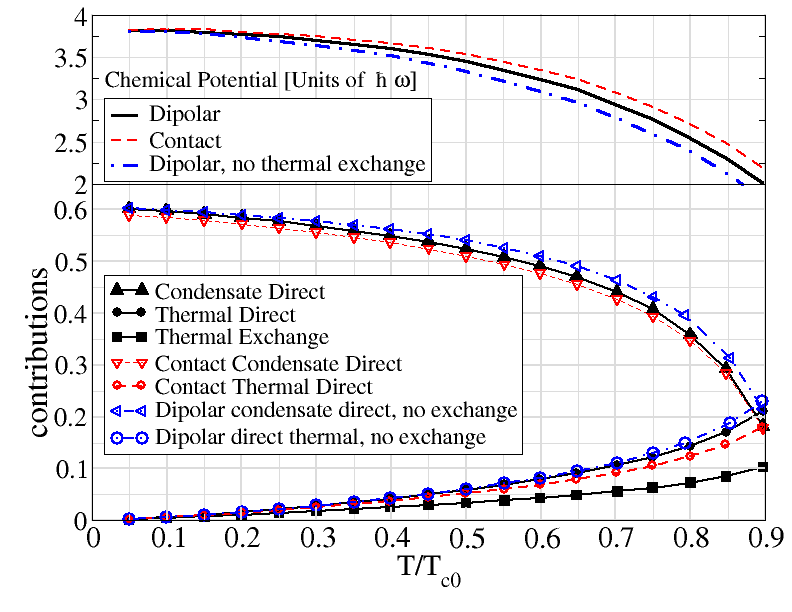}
\caption{The chemical potential for three systems: dipolar (black), contact (dashed red), 
and dipolar without exchange (dash dot blue). The interaction terms for each system in 
the Hamiltonian as a function of temperature. The contributing terms are
the direct condensate ($\langle \phi_0 |D_0|\phi_0\rangle/\mu$, triangles), 
the direct thermal ($\langle \phi_0 |\tilde D|\phi_0\rangle/\mu$, circles), 
and the thermal exchange ($\langle \phi_0 |\tilde X/|\phi_0\rangle/\mu$, squares) 
interaction. }\label{chemcomp}
\end{figure}

In Fig. \ref{chemcomp} we look at the chemical potential and its
contributions from Eq. \ref{GPE}, as a function of temperature.  
The examples shown are dipolar (black), contact (red), and
dipolar without thermal exchange (blue).
We show the direct condensate interaction ($\langle \phi_0 |D_0|\phi_0\rangle/\mu$, 
triangle), the direct thermal interaction ($\langle \phi_0 |\tilde D|\phi_0\rangle/\mu$, 
circle), and the  thermal exchange interaction 
($\langle \phi_0 |\tilde X/|\phi_0\rangle/\mu$, square). 
For the contact interaction, the thermal exchange is equal to thermal direct
interaction, so  we only show one. 
For each example, the remaining contribution to the chemical 
potential is from $H_0$ or the potential and kinetic energy contributions.

For the dipolar gas, the thermal exchange term is about half as strong the 
direct contribution.
The importance of the interaction with the thermal particles is more important
as the temperature is increased. 
Comparing the dipolar calculation with and without the thermal exchange,
we see that the chemical potentials agree reasonably well, the full calculation
has a larger $\mu$ at high temperature.
Looking at the interaction contributions, we see that the direct condensate 
(black triangle) and thermal interaction (black circles)
make up a smaller portion of the chemical potential for 
the full calculation than the one with $\tilde X=0$ 
(blue triangle and open circles).

\begin{figure*}
\includegraphics[width=180mm]{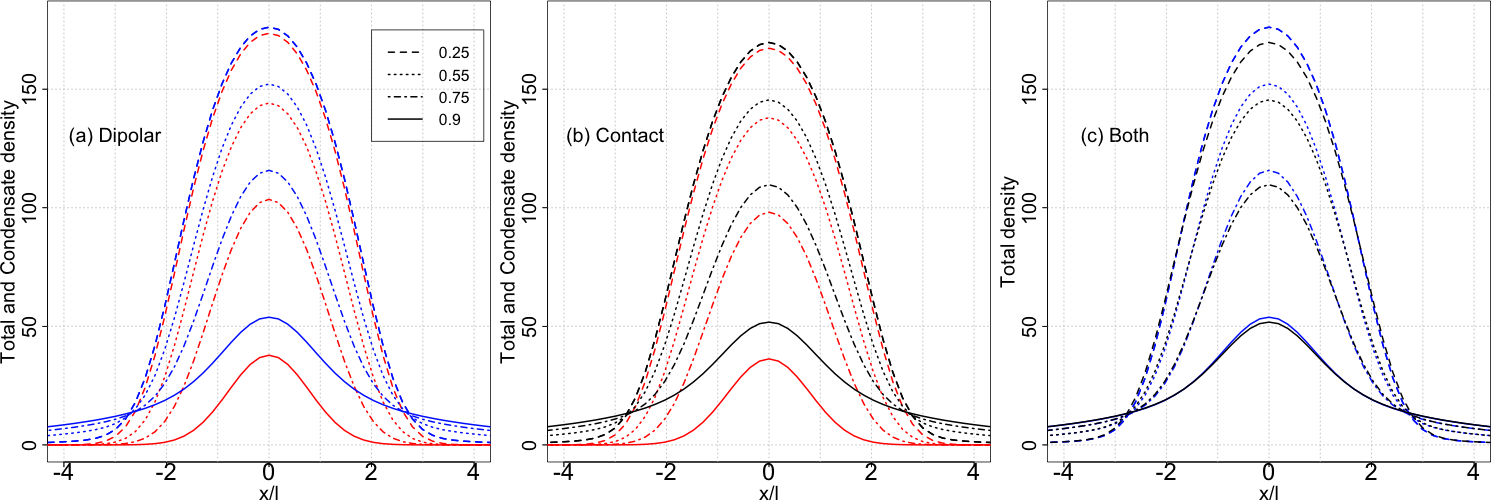}
\caption{The total and condensate density of (a) a Cr dipolar gas and 
(b) contact gas, and the comparison of the total densities (c). (a) The 
total density is in blue (black) for the dipolar (contact) gas, and the 
condensate density is in red. The temperatures from top to bottom are 
$T/T_{c0}=$ 0.25 (dashed), 0.55 (dotted), 0.75 (dash-dot), and 0.90 (solid).
(c) The dipolar gas is more dense in the center of the trap at all 
temperatures.}\label{denCOMP}
\end{figure*}
In Fig. \ref{denCOMP} we show both the total and condensate density 
for (a) a Cr system of 2000 particles and an analogous (b) contact system 
(equal chemical potentials at $T=0$).  
We then compare the density profiles in (c). The total density is in blue 
(black) for the dipolar (contact) systems, 
and the condensate density is shown in red.  The temperatures from top to 
bottom are 
$T/T_{c0}=$ 0.25 (dashed), 0.55 (dotted), 0.75 (dash-dot),  and 0.90 (solid). 
For both the contact and dipolar gas at large $\rho$, the thermal density 
behaves as the analytic solution predicts.  

An important point is that the dipolar gas has a higher density 
in the middle of the trap than the contact gas.
It is hard to see in the figure, but the contact condensate atoms 
have been shifted to the shoulders of the condensate or near the trap edge.  
This has implications 
for the temperature at which the superfluid phase transition will occur in 
a dipolar gas.  

\begin{figure*}
\includegraphics[width=180mm]{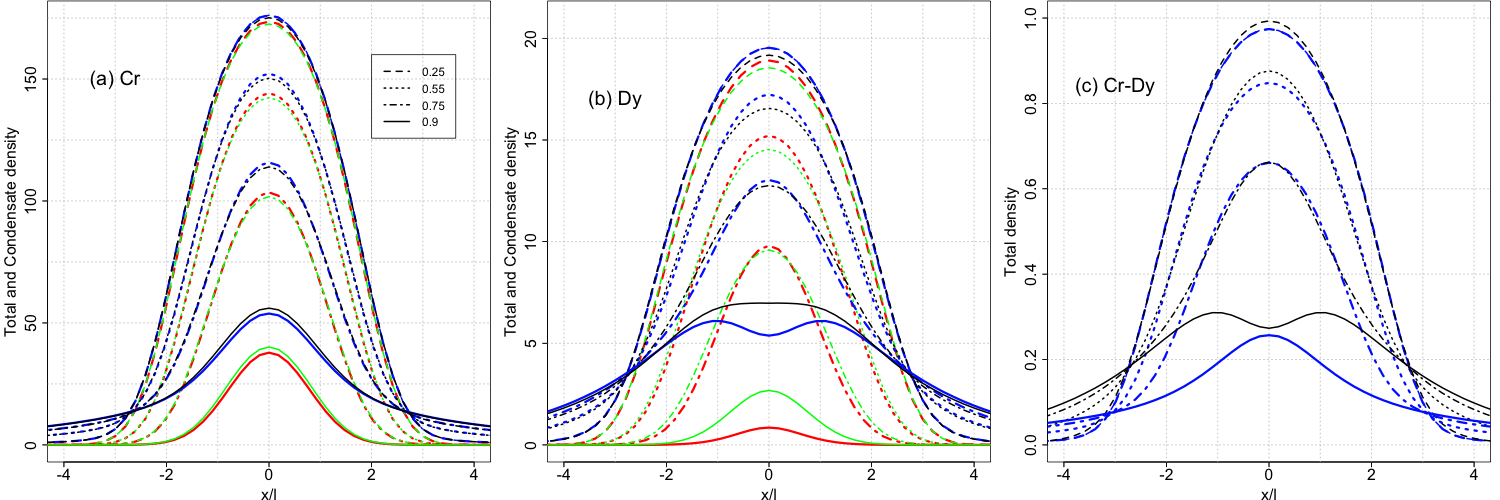}
\caption{Comparison of density with and without thermal exchange. 
Both the total and condensate density are shown at four 
temperatures, top to bottom:
$T/T_{c0}=$ 0.25 (dashed), 0.55 (dotted), 0.75 (dash-dot), and 0.90 (solid).
(a) A Cr gas with (blue) and without (black) the thermal exchange;
the corresponding condensate density is with (red) and without (green) 
thermal exchange is shown.
(b) For a Dy gas, the total density with (blue) and without (black) the thermal 
exchange; the corresponding condensate density with (red) and without (green) 
thermal exchange is shown.  Both the condensate and total density
without thermal exchange have a lower central density.
(c) Shows a comparison of Dy (black) and Cr (blue) with a large number of atoms (3700)
when the chemical potential are the same at zero temperature.
The densities are normalized by their $T=0$ central density.
}\label{denCOMP2}
\end{figure*}

In Fig. \ref{denCOMP2} we compare both the total and condensate density 
with and without the thermal exchange.
We look at four temperatures, from top to bottom they are:
$T/T_{c0}=$ 0.25 (dashed), 0.55 (dotted), 0.75 (dash-dot), and 0.90 (solid).
In Fig. \ref{denCOMP2}  (a), a Cr gas with (blue) and without (black) the thermal exchange is 
shown.  Additionally, the corresponding condensate density is with (red) 
and without (green) thermal exchange.
This example with the thermal exchange is the same as Fig. \ref{denCOMP} (a). 
In Fig. \ref{denCOMP2}  (b) a Dy gas with (blue) and without (black) the 
thermal exchange is shown.
The corresponding condensate density is shown with (red) and without 
(green) thermal exchange.
\ref{denCOMP2}  (c) Shows a comparison of Dy and Cr with a large number of atoms (3700),
when the chemical potentials are nearly the same at zero temperature.
The densities are normalized to the maximum density at zero temperature.

In Fig. \ref{denCOMP2}  (a) the impact of the thermal exchange is not too pronounced. 
It slightly increases the condensate central density, which is most obvious at
high temperature. This is significantly different
from the Dy example shown in Fig. \ref{denCOMP2}  (b).  There is a noticeable difference 
between the density 
profiles with and without the thermal exchange.  The effect is most clearly seen
in the condensate by looking at the increased central density and reduced extent
of the red curve (with) versus green (without).
At the highest temperature, where the validity of HFBP is questionable, we see a 
significant reduction in the condensate density and occupation.
In fact the condensate repels the thermal cloud so strongly, that in the center of 
the trap there is a local minimum in the total density at $T/T_{c0}=0.9$
Furthermore, the thermal exchange has significantly lowered the
condenstate occupation.  Thus the inclusion of the thermal exchange
clearly lowers the critical temperature of the dipolar gas. 

In Fig. \ref{denCOMP2}  (c) we show that by varying $g_d$ and $N$ while holding
$Ng_d$ constant, the HFBP does not give identical results as the $T=0$ GPE would.
For $g_d=0.034, N=300$ (Dy, black) and $g_d=0.0028,N=3700$ (Cr, blue) 
at low temperatures the density profiles are slightly different. 
The densities are normalized by their $T=0$ central density.
More importantly, at high temperature the profiles are very different, this has 
to do with the strong depletion of the condensate. 

\begin{figure}
\includegraphics[width=80mm]{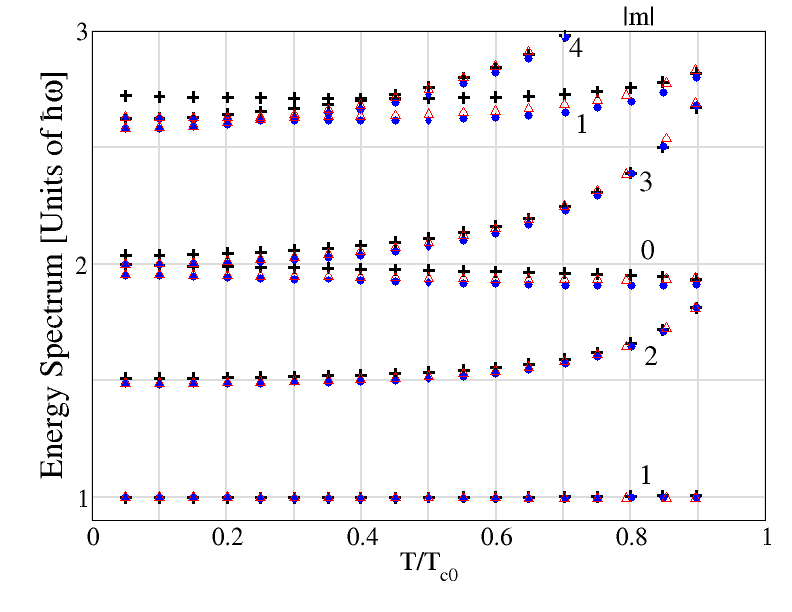}
\caption{The excitation spectrum of the dipolar (blue circles),
dipolar without exchange (red triangle), and
 contact gas (black +) as a function of temperature for $N=2000$.
The azimuthal symmetry of the excitations are shown next to each curve.}
\label{bdgFIG}
\end{figure}
In Fig. \ref{bdgFIG} we show the excitation spectrum of the dipolar gas 
(blue circles), a dipolar gas without the thermal exchange (red triangle),
and the contact gas (black +) as a function of temperature.
We have also characterized the excitations by their azimuthal symmetry. 
In a 2D contact gas, there is a hidden symmetry which makes the breathing 
mode ($m$=0) have an energy of 2$\hbar\omega$ \cite{pit}.  This mode has very little 
temperature dependence.  This hidden symmetry is removed by the dipolar interaction, 
however if $l_z$ goes to zero, the breathing mode goes to 2$\hbar\omega$.
In the example we have picked ($g_d=0.0028$ and $N$=2000), the $|m|=3$ mode is
near 2$\hbar\omega$, but this is not the breathing mode ($m=0$).  
Rather, the mode just below this, is the breathing mode. 
The contact gas excitation spectrum (black +) agrees well with 
Ref. \cite{gies}. They used $g\sim0.002$ which accounts 
for the differences.  It is a general feature that the dipolar gas has lower excitation
energies than the contact gas. In Ref. \cite{ronen}, the 3D excitation spectrum as a 
function of temperature showed that for an oblate geometry 
the dipolar excitations are lower than the contact system.
It is worth pointing, out that the dipolar calculations with and without the thermal
exchange, agree well at low temperature.  As the temperature is increased there is
some disagreement between the two (blue circle and red triangle), but mostly for more 
highly excitated modes.  The calculations without the exchange become higher in energy.

The modes that have strong temperature dependence are those with higher 
azimuthal symmetry.  In contrast, the full bodied modes (large central amplitude)
have a more constant excitation frequency as temperature is varied.  
The reason for this is discussed below. 

\begin{figure*}
\includegraphics[width=75mm]{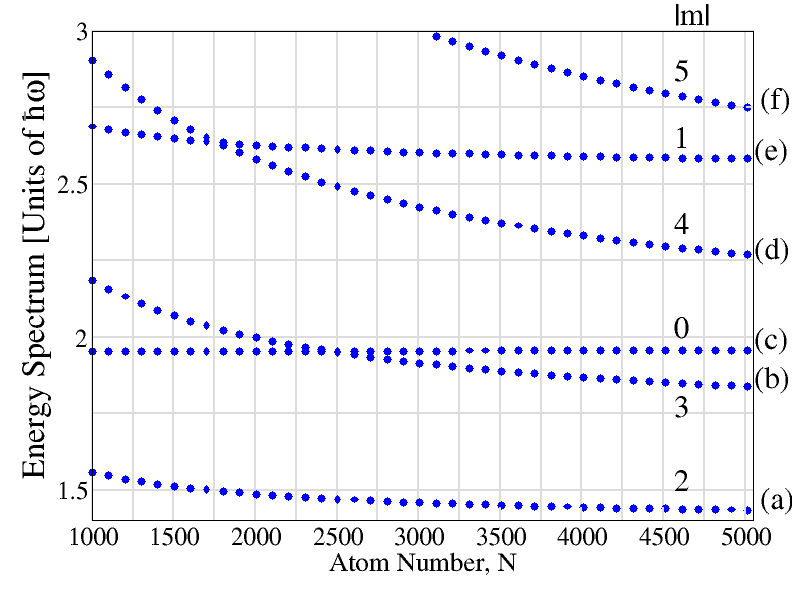}
\includegraphics[width=94mm]{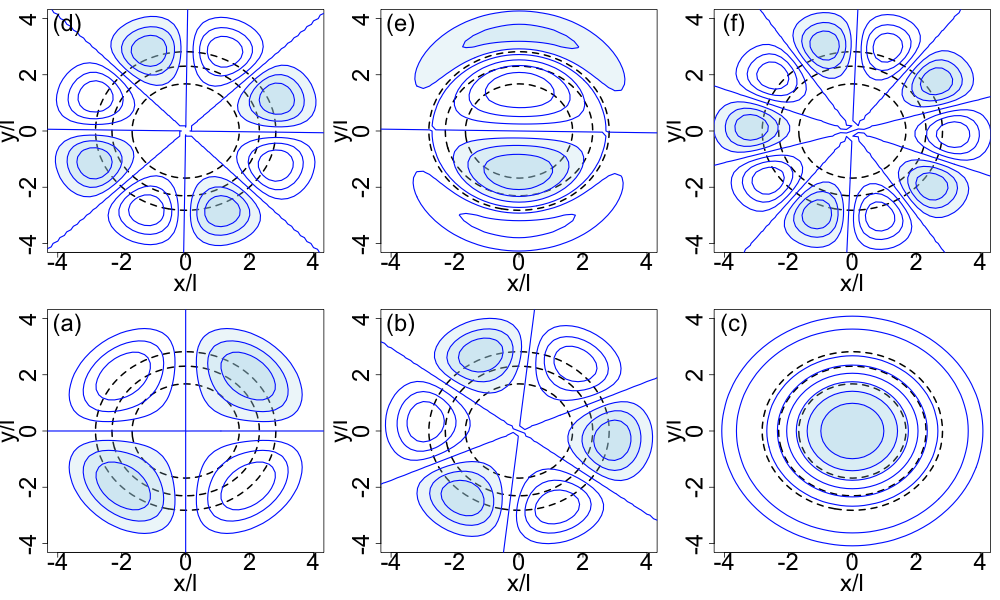}
\caption{The excitation spectrum as a function of $N$ for the dipolar 
(blue circles) at $T/T_{c0}=0.05$.
The plots on the right are contour plots of the quasi-particle, $u_\alpha$, 
modes for $N=5000$ dipolar atoms. The contour lines are are in 0.25 
increments of the maximum value of $u_\alpha$ (i.e., 0.75, 0.5, ...,-0.75). Additionally, 
the negative regions are shaded blue.  The condensate density contours (dashed black) are on
the same scale.}
\label{bdgFIG2}
\end{figure*}

In Fig. \ref{bdgFIG2} we show the excitation spectrum as a function of $N$ 
for the dipolar (blue circles) at $T/T_{c0}=0.05$.
The plots on the right are contour plots of the quasi-particle modes ($u_\alpha$)
for $N=5000$ dipolar atoms. The contour lines are in 0.25 increments of the maximum 
value of $u_\alpha$
(i.e., 0.75, 0.5, ...,-0.75). Additionally, the negative regions are shaded 
blue.  The condensate density contours are shown as a dashed black lines.
The azimuthal symmetries are: (a) $m$=2 quadruple mode, (b) $m$=3, (c) $m$=0, 
breathing mode, (d) $m$=4, (e) $m$=1, and (g) $m$=5.  
We have shown one of the two degenerate modes ($\pm m$) at each energy 
when $m\ne0$.
If the contact system were plotted, we would see that excitations are always
slightly higher in energy. 

The modes that are not strongly dependent on number or temperature are the 
modes that have a full bodied motion or large amplitude in the middle of the 
condensate, see \ref{bdgFIG2} (c) and (e). These modes are mildly impacted by
the total number of atoms in the system. 
The modes with higher azimuthal symmetries are more like surface modes, and  are 
strongly effected by the number or temperature,  see \ref{bdgFIG2} (a), (b), (d), 
and (f). 

In fact the strong number dependence leads to excitations crossing paths as 
$N$ is increased.  The $m$=3, (b), mode becomes lower than the breathing mode, (c); 
and $m$=4, (d), mode becomes lower than the $m$=1, (e). The fact that surface modes 
move relative to full body excitations as a function of $N$ is related to the size 
of the condensate. Both as the atom number is increased, and as the temperature is 
decreased, the size of the condensate becomes larger.
Naively, one might consider the surface of the condensate a ring with a restorating
force.  We consider the surface excitations as displacements of the ring 
from equilibrium, and find an excitation spectrum which behaves as $m/R$ where 
$R$ is radius of the ring \cite{fetter2}. 
Then we look at $m$ over the radius at which the density equals $0.25n_0(0)$
(the furthest out contour of the condensate in Fig. \ref{bdgFIG2}), 
we find a similar number dependence to the excitation spectra 
in Fig. \ref{bdgFIG2}.  This is too simplistic, but gives a physical reason 
for such behavior in the excitation spectrum.
The important point of this analysis is that the size of the condensate grows
quickly at small $N$ and slows down at larger $N$.  Thus we see a rapid change
in the excitation spectrum at low $N$ and less so at large $N$.

\section{conclusion}
This paper studied a trapped q2D dipolar gases at 
finite temperature.  We presented the numerical method used to solve
Hartree Fock Bogoliubov Popov equations
for a gas with non-local interactions including thermal exchange effects.  
In Fig. \ref{ncomp} we showed condensate fraction as a function of
temperatue for both the Cr and Dy examples. 
The critical temperature of the Dy is greatly reduced by the dipolar interaction.  
In Fig. \ref{chemcomp} we looked at the chemical potential and the
total interaction energy as a function of temperature 
for the various terms in Eq. \ref{GPE}.
We compared the contact example with the Cr
examle with and without thermal exchange. 
In Fig. \ref{denCOMP} we studied the impact of temperature on the
density profiles and saw that the dipolar gas is more dense in the 
center of the trap. 

Next in  Fig. \ref{denCOMP2} we explored the impact of the thermal
exchange on both the Cr and Dy examples.  We found that the strongly
interaction Dy example, the thermal exchange strongly reduces the
condensate fraction at high temperature and alters the density profile
by making it more narrow and dense in the center of the trap. 
In Fig. \ref{bdgFIG} we studied the impact of temperature on the
excitation spectra.  We compared the contact example with the Cr
examle with and without thermal exchange. 
Finally, in Fig. \ref{bdgFIG2} we looked at the excitation spectrum of
a Cr gas as a funciton of atom number and the quasi-particles,
$u_\alpha$, for $N=5000$. Here we illustrated the strong
number or size dependence of the high azimuthal symmetries.

This work has set the stage to study phase coherence of a dipolar gas
as a function of temperature, as has been done for contact gases \cite{gies}.
We seek to understand how the non-local interaction will alter the correlations
at the BKT transition.  We already saw an increased 
phase space density in the middle of the cloud from the dipolar 
interaction. How does the dipolar interaction impact the phase coherence 
of the gas?
Second, we wish to study roton physics in q2D \cite{roton,Wilson,Wilson2}. 
In q2D, a roton can emerge as the field is tilted into the plane of motion, and
leads to anisotropic density profiles \cite{aniso}.  The presented numerical
method has been developed to handle this configuration, and lead to experimentally
observable signatures of the roton. 

This method could be applied to other momentum dependent interactions,
for instance those with renormalized contact interactions, such as 
$g\rightarrow(g+g^2/k^2)$ \cite{fetter}.  
Additionally, we could improve the scattering model to include
more dipolar scattering physics \cite{CT2d}.  
\begin{acknowledgments}
The author greatfully acknowledges support from the
Advanced Simulation and Computing Program (ASC) and LANL which is operated 
by LANS, LLC for the NNSA of the U.S. DOE under Contract No. DE-AC52-06NA25396.
\end{acknowledgments}
\bibliographystyle{amsplain}

\end{document}